# Spintronic nano-scale harvester of broadband microwave energy

Bin Fang[1], Mario Carpentieri[2], Steven Louis[3], Vasyl Tiberkevich[4], Andrei Slavin[4], Ilya N. Krivorotov[5], Riccardo Tomasello[6], Anna Giordano[6], Hongwen Jiang[7], Jialin Cai[1], Yaming Fan[1], Zehong Zhang[1], Baoshun Zhang[1], Jordan A. Katine[8], Kang L. Wang[9], Pedram Khalili Amiri[9,10*], Giovanni Finocchio[6*], and Zhongming Zeng[1*]

[1]Key Laboratory of Nanodevices and Applications, Suzhou Institute of Nano-tech and Nano-bionics, Chinese Academy of Sciences, Ruoshui Road 398, Suzhou 215123, P. R. China

[2]Department of Electrical and Information Engineering, Polytechnic of Bari, Bari 70125, Italy

[3]Department of Electrical and Computer Engineering, Oakland University, Rochester, Michigan 48309, USA

[4]Department of Physics, Oakland University, Rochester, Michigan 48309, USA

[5]Department of Physics and Astronomy, University of California, Irvine, California 92697, USA

[6]Department of Mathematical and Computer Sciences, Physical Sciences and Earth Sciences, University of Messina, Messina 98166, Italy

[7]Department of Physics and Astronomy, University of California, Los Angeles, California 90095, USA

[8]Western Digital, San Jose, California 95138, USA

[9]Department of Electrical and Computer Engineering, University of California, Los Angeles, California 90095, USA

[10]Department of Electrical Engineering and Computer Science, Northwestern University, Evanston, Illinois 60208, USA

# Abstract

The harvesting of ambient radio-frequency (RF) energy is an attractive and clean way to realize the idea of self-powered electronics. Here we present a design for a microwave energy harvester based on a nanoscale spintronic diode (NSD). This diode contains a magnetic tunnel junction with a *canted magnetization of the free layer*, and can convert RF energy over the frequency range from 100 MHz to 1.2 GHz into DC electric voltage. An attractive property of the developed NSD is the generation of an almost constant DC voltage in a wide range of frequencies of the external RF signals. We further show that the developed NSD provides sufficient DC voltage to power a low-power nanodevice - a black phosphorus photo-sensor. Our results demonstrate that the developed NSD could pave the way for using spintronic detectors as building blocks for self-powered nano-systems, such as implantable biomedical devices, wireless sensors, and portable electronics.

Energy harvesting technologies offer a promising approach to capture energy from ambient sources, such as vibration, heat, and electromagnetic waves. Among them, the ambient radio-frequency (RF) electromagnetic signals provide an attractive energy source for applications in self-powered portable electronics in the "internet of things" era, where device miniaturization is often constrained by the battery technology[1]. Electromagnetic devices used for microwave energy harvesting could be efficient, but are difficult to miniaturize[2], while semiconductor devices, such as Schottky diodes, fail to provide a satisfactory conversion efficiency specifically for micro-Watt power-harvesting, mainly because of their high zero-bias junction resistance[3]. Spintronic technologies, which use both charge and spin of electrons, could become a viable alternative for the development of novel energy-efficient broadband electronic systems (e.g. magnetic sensors, memory, oscillators and detectors[4-9]), and also for the design of miniature energy harvesters, such as thermoelectric power generators[10,11]. In particular, it has recently been shown theoretically, that spintronic diodes could also play the role of RF energy harvesters[3,12]. Below, we report the development of a *bias-free* nanoscale spintronic diode (NSD) based on a magnetic tunnel junction (MTJ) having a *canted magnetization* in the free layer, and demonstrate that this NSD could be an efficient harvester of broadband ambient RF radiation capable of powering a low-power semiconductor nanodevice - a black phosphorus photo-sensor.

The operational principle of NSDs used as electromagnetic energy harvesters is illustrated in Fig. 1a. Typically, a weak *ac* current will be spin-polarized as it passes through a fixed magnetic layer and enters the free layer of a MTJ. This spin-polarized current excites a small-amplitude magnetization precession (Fig. 1a, trajectory in red) which then creates a rectification voltage $V_{DC}$ across the MTJ that is highest near the ferromagnetic resonance (FMR) frequency[9,13-17]. Large sensitivities were observed under application of a DC voltage and/or external magnetic field. However, this requires energies larger than the ones harvested[18,19]. Moreover, broadband detection is essential for energy harvesting from ambient RF and microwave energy sources since their exact frequencies are, generally, unknown. Thus, resonant spintronic diodes are not suitable for RF energy harvesting. The novel MTJ design in this study exhibits large-amplitude magnetization precession (Fig. 1a, trajectory in blue) excited by a microwave current that is relatively weak, which allows one to substantially increase the rectification voltage $V_{DC}$ over a broadband frequency range. This result is accomplished by fine-tuning the magnetic anisotropy[20,21] to create a free layer with the equilibrium magnetization directed along an oblique easy axis. Experimental data and micromagnetic simulations reveal that this *oblique* design is the key ingredient that enables the excitation of a large-amplitude magnetization precession without bias current and without bias magnetic field, over a broad frequency range and at room temperature, hence allowing to accumulate energy from external signals of several different frequencies acting simultaneously, and thereby delivering output power levels relevant for energy harvesting scenarios.

## Results

**NSD device.** The NSD energy harvester based on an MTJ was fabricated with a continuous multilayer thin film stack composed of PtMn (15)/$Co_{70}Fe_{30}$ (2.3)/Ru (0.85)/$Co_{40}Fe_{40}B_{20}$ (2.4)/MgO (0.8)/$Co_{20}Fe_{60}B_{20}$ (1.65) (thickness in nm) (Fig. 1b). The free layer is composed of $Co_{20}Fe_{60}B_{20}$, while the MgO tunnel barrier ensures a high magnetoresistance effect[22,23]. The perpendicular magnetic anisotropy (PMA) in the $Co_{20}Fe_{60}B_{20}$/MgO bilayer is caused by an interfacial effect that arises from the hybridization between the O (of the MgO) and Fe (of the CoFeB) orbitals, and can be controlled by the CoFeB composition and thickness of the free layer[20,21,24]. In other words, below a critical thickness the free layer easy axis is out-of-plane. The devices studied here are designed to have a free layer thickness near yet below this critical value, so that the PMA field almost compensates for the out-of-plane demagnetizing field ($K_1 - 0.5\mu_0 M_S^2 \approx 0$, where $K_1$ is the first order anisotropy coefficient, $\mu_0$ is the vacuum permeability, and $M_S$ is the saturation magnetization). In this scenario, the second-order magnetic anisotropy ($K_2$ being the second order anisotropy coefficient) plays a crucial role in controlling the magnetization configuration, even though $K_2$ is smaller than $K_1$ by about one order of magnitude[25]. The magnetic multilayer film stack was patterned into elliptical nanopillars by using electron-beam lithography and ion milling techniques. The representative results presented here are from a device with dimensions of 150 nm $\times$ 50 nm, while similar results have been obtained on a variety of samples with other dimensions. All data were gathered at room temperature.

We first characterized the resistance of the NSD as a function of the magnetic field applied parallel to the ellipse major axis ($H_{//}$) or perpendicular to the sample plane ($H_{\perp}$) (see one example in Supplementary Fig. 1), which confirms that a perpendicular anisotropy exists in the free layer, while the gradual approach to saturation for both field directions indicates the free-layer magnetization has a canted equilibrium state in the absence of an external field. For the device presented here, the tunnel magnetoresistance (TMR) ratio, defined as $(R_{AP}-R_P)/R_P$, is 93%, and the equilibrium angle ($\theta$) at zero bias field is estimated to be 53°, calculated from the field scans of the magnetoresistance as[18]

$$R^{-1}(\theta) = \frac{R_P^{-1} + R_{AP}^{-1}}{2} + \frac{R_P^{-1} - R_{AP}^{-1}}{2}\cos(\theta), \tag{1}$$

where the resistances in the parallel ($R_P$) and antiparallel ($R_{AP}$) configurations are 640 Ω and 1236 Ω, respectively.

**Rectifying properties.** Next, we measured the rectified voltage with an RF current in the absence of an applied magnetic field. As shown in Fig. 1b, the RF current $I_{ac}\sin(2\pi f_{ac}t)$ is applied to the device through a bias tee using a signal generator (N5183B, Keysight), while the rectification voltage $V_{DC}$ across the MTJ is recorded with a nanovoltmeter (K2182, Keithley). The frequency response of the NSD exhibits two qualitatively different behaviors:

a. ***Resonant detection***: For a relatively low incident RF power, $P_{RF} = 0.1$ µW, a resonant type response is observed with a maximum near the FMR frequency of 1.1 GHz (Fig. 2a). The line-shape associated with this peak in voltage can be fitted by a sum of symmetric and anti-symmetric Lorentzian functions, as discussed in previous studies[14-16,26,27]. The origin of the asymmetric line shape is related to the voltage-controlled interfacial perpendicular magnetic anisotropy (VCMA) effect[16].

b. ***Broadband energy harvesting***: For larger input RF power, a novel type of broadband response is achieved. Figure 2b displays an example with a RF input power of $P_{RF} = 10$ µW, showing that the NSD rectifies a nearly constant voltage across a 100 to 550 MHz range.

Other fabricated NSDs showed qualitatively similar results with different bandwidths and output voltages, with the widest harvesting RF energy across a 1.2 GHz bandwidth (Supplementary Fig. 2 and Supplementary Note 1).

The transition from resonant detection to broadband energy harvesting can be understood in terms of the evolution of the precession orbit[12]. At low $P_{RF}$, a small-amplitude magnetization precession is excited where the magnetization oscillates in a circular trajectory as shown in Fig. 2c (or Supplementary Video 1). When $P_{RF}$ is sufficiently large, the free layer oscillation becomes a large-amplitude magnetization precession (Fig. 2d, or Supplementary Video 2), and as such results in a large rectified voltage. In the broadband region, a change of input frequency does not change the magnetization precession amplitude in the $x$ direction, and thus the rectification voltage remains nearly independent of the input frequency (Fig. 2b). Similarly, the rectification voltages are approximately independent of the $P_{RF}$ value after it exceeds a threshold[12], because the $x$-component of the magnetization oscillates near the maximum amplitude (see the projection in the $x$-$y$ plane in Fig. 2d), whereas additional input power is then converted to the excitation of higher non-linear modes and does not contribute to the rectified voltage.

Figure 3 shows the experimental performance of the device at two frequencies, 250 MHz and 500 MHz, with input RF power varied across a considerable energy range. It is evident that when the input power is above a threshold of about 2 μW, the non-resonant character is achieved and the generated voltages are almost constant for all the measured values of $P_{RF}$. This behaviour differs qualitatively from the resonant detection in spintronic diodes[9,16,18,19], where the generated voltage is strongly reduced when the incoming RF power is reduced. The experimentally observed behavior is well reproduced by micro-magnetic computations, shown in Fig. 3. It is worth noting that Jenkins A. *et al*[28] reported analogous behavior in spin-torque resonant expulsion of the vortex core under a large DC bias and magnetic field. While their scheme could be used as a tunable microwave detector, the very large power input required for biasing makes it not suitable for energy harvesting. Thus, in the broadband scheme, the NSD operates as a non-resonant broadband microwave energy harvester for low-frequency RF signals.

**Advantage of broadband detection for Energy harvesting.** Last, we have performed measurements to demonstrate that an individual NSD nanogenerator can work as a power source for a real world circuit. Figure 4a shows the measurement configuration. The NSD energy harvester is connected to the drain of a high-performance few-layer black phosphorus (BP) nanodevice, chosen for its low power operating capability[29,30]. The NSD supplies the voltage ($V_{ds}$) by converting energy that it receives from RF sources through an antenna. Three different RF sources are used in the experiment: (i) Signal generator, (ii) Walkie talkie, and (iii) Radiophone. Figure 4b shows current $I_{ds}$ as a function of $V_{ds}$ generated under RF energy from signal generator. The linear $I_{ds}$-$V_{ds}$ curve achieved proves that NSD can be used to bias the nanodevice. Our results are now extended to consider different sources that can give rise to ambient background RF energy, we have found that a rectified DC voltage is observed when a Walkie talkie (462 MHz) or a Radiophone (923 MHz) is turned on near the antenna connected to the NSD. As an example, as displayed in Figure 4c, for a 6.3 µW (Walkie talkie) and a 4.0 µW (Radiophone) input RF power received through the antenna, a voltage of 150 µV and 100 µV is measured, respectively.

A key benefit of the NSD with capability of broadband detection is that it provides a possible solution to generate larger DC voltages when multiple external RF sources are added to the NSD device simultaneously. In order to verify this, we conducted the measurements under two RF sources with different frequencies. Figure 4d shows an example of the rectified DC voltages under two RF sources, exhibiting that the rectified DC voltage is larger when two external RF sources of different frequency are acting on the NSD device simultaneously, compared to the case when only one RF source is present. This makes it possible to generate more harvested power than shown in this demonstration, if multiple simultaneous external signals act on the NSD device. The measurements for different RF sources clearly demonstrate that the developed NSD can be used as an efficient harvester of the broadband ambient RF energy.

Furthermore, we consider an example of how an NSD can be used as a power supply for a BP device acting as a photo-sensor. The NSD provides bias voltage while the BP acts as a sensor of photons, and can be triggered by a typical light source. Figure 4e shows the response of this BP sensor to a modulated light when the sensor is biased by a voltage of 100 µV, produced by the NSD that is harvesting the Radiophone RF energy: when the light turns on, the BP enters the on-state, and the resistance drops, allowing $I_{ds}$ to increase to 5 nA. When the light turns off, the BP enters the off-state and the resistance increases, lowering $I_{ds}$ to 1 nA. This phenomenon is consistent with a previous report on BP photo-detection[31]. These experimental data prove that an NSD harvesting ambient RF signals can be used to bias nanodevices, thus opening a path for a possible implementation of various applications of self-powered devices.

## Discussion

The conversion efficiency is one of the key parameters to characterize the performance of the energy harvester for practical applications[32]. The overall conversion efficiency is determined by a combination of several factors: the matching efficiency, the parasitic component efficiency, the RF-to-DC conversion efficiency and the DC power transfer efficiency[3]. For an RF diode, it is reasonable to use the RF-to-DC conversion efficiency $\eta_C$ as the most relevant figure of merit, since the other parameters primarily relate to the circuit and component design rather than the underlying physics of energy conversion. The RF-to-DC conversion efficiency $\eta_C$ is defined as the generated DC power ($P_{DC}$) divided by the input RF power ($P_{RF}$), in which $P_{DC} \approx V_{DC}^2/R_{MTJ}$, where $R_{MTJ}$ (708 Ω in this study) is the zero-bias resistance of the spintronic diode at zero magnetic field.

For an input power of $P_{RF} = 3.2$ μW, the spintronic diode measured in this work has an efficiency of 0.005%. This value is small compared to the low-barrier Schottky diodes with low-power capability[33] (0.1~0.4%, measured under the same conditions, see Supplementary Figs. 3-5 and Supplementary Note 2). However, it is important to note that the generated DC voltage in this work is a fraction of the expected maximum value ($V_{DC} < 10\%\ V_{DC}^{max}$) (see Supplementary Note 2), which promises a large room for future improvement. Furthermore, an improved diode design with larger TMR value should lead to a significant improvement of the efficiency. Large TMR ratios of > 200% have been already achieved by several groups, but these MTJ stacks, typically, are optimized for magnetic memory applications. For example, Ikeda *et al* reported a TMR ratio of 604% in an in-plane magnetized MTJ structure CoFeB/MgO/CoFeB at room temperature[34]. Almasi *et al.* obtained a TMR as high as 208% in an MTJ with perpendicular magnetic anisotropy by optimizing the Mo dusting thickness[35]. Kubota *et al.* theoretically predicted a TMR of 600% for the $D0_{22}$-$Mn_3Ga$/MgO/$Mn_3Ga$ (001) MTJ system[36]. In addition, the total efficiency can be also improved by optimization of antennas, tunnel barrier (enhancing the nonlinearity and consequently, the efficiency), and impedance matching[37,38]. Therefore, following theoretical predictions[12], spintronic diodes with optimized structures are expected to surpass the Schottky diodes for low-power energy harvesting.

In summary, we have experimentally demonstrated that spintronic diodes based on MTJs with canted free layers can act as nanoscale electromagnetic energy harvesters, opening a new direction for research in nano-scale spintronic energy harvesters. Our results also demonstrate that the power generated by NSDs is sufficient to drive active devices, such as BP nano-sized photo sensors, for use in low-power electronic systems, showing their potential as building blocks of self-powering devices for applications in wireless sensing and portable electronics in the emerging era of the "internet of things".

## Methods

**Sample preparation.** The continuous multilayer thin films with stacks of composition PtMn (15) /Co$_{70}$Fe$_{30}$ (2.3)/Ru (0.85)/Co$_{40}$Fe$_{40}$B$_{20}$ (2.4)/MgO (0.8)/Co$_{20}$Fe$_{60}$B$_{20}$ (1.65) (thickness in nm) were deposited using a Singulus TIMARIS physical vapor deposition system and annealed at 300 °C for two hours in a magnetic field of 1 T. The films were subsequently patterned into ellipse-shaped pillars using electron-beam lithography and ion milling techniques.

**Micromagnetic simulations.** We numerically solved the Landau-Lifshitz-Gilbert-Slonczewski equation which includes the field-like torque $T_{OP}$ (ref. 39), and the voltage dependence of the anisotropy, i.e. VCMA[16]. The $T_{OP}$ is considered to be dependent on the square of the bias voltage up to a maximum value of 10% of the in-plane torque computed for a current density |$J$| 4.0×10$^6$A/cm$^2$. The total torque, including also the in-plane component $T_{IP}$ is given by

$$T_{IP} + T_{OP} = \frac{g\,|\mu_B|\,J}{|e|\,\gamma_0\,M_s^2\,t}\,g_T(\mathbf{m},\mathbf{m_p})\,\left[\mathbf{m}\times(\mathbf{m}\times\mathbf{m_p}) - q(V)(\mathbf{m}\times\mathbf{m_p})\right] \qquad (3)$$

where **m** and **m$_p$** are the normalized magnetizations of the free and pinned layer respectively, $g$ is the gyromagnetic splitting factor, $\gamma_0$ is the gyromagnetic ratio, $\mu_B$ is the Bohr magneton, $q(V)$ is the voltage-dependent parameter for the perpendicular torque, $J$ is the current density, $V$ is the voltage, $t$ is the thickness of the free layer, and $e$ is the electron charge. **m$_p$** is considered fixed along -$x$ direction. The effective field takes into account the standard micromagnetic contributions (exchange, self-magnetostatic) as well as the dipolar coupling from the pinned layer, the Oersted field due to both microwave and d.c. current and both first and second order PMA $h_{ani-z} = \frac{2K_1 m_z + 4K_2 m_z(1-m_z^2)}{\mu_0 M_S^2}$ ($h_{ani-x} = h_{ani-y} = 0$) where $m_z$ is the normalized z-component of the magnetization. The presence of the VCMA has been modeled as described in ref. 40. The parameters used for the CoFeB are saturation magnetization $M_s = 9.5 \times 10^5$ Am$^{-1}$, exchange constant $A = 2.0 \times 10^{-11}$ Jm$^{-1}$, and damping parameter $\alpha = 0.02$. The $K_1 = 5.52 \times 10^5$ Jm$^{-3}$ and $K_2 = 1.5 \times 10^4$ Jm$^{-3}$ have been estimated from fitting of the resonant frequency versus magnetic field[16] (Supplementary Fig. 6 and Supplementary Note 3), while the VCMA coefficient is 35 fJV$^{-1}$m$^{-1}$ (Supplementary Fig. 7)[16]. We used the spin polarization function $g_T(\mathbf{m},\mathbf{m_p}) = 2\eta_T(1+\eta_T^2 \mathbf{m}\cdot\mathbf{m_p})^{-1}$ computed by Slonczewski[41, 42] with a spin-polarization $\eta_T = 0.6$[13].

**Acknowledgements**

This work was supported by the National Science Foundation of China (11474311) and the executive programme of scientific and technological cooperation between Italy and China for the years 2016–2018 (2016YFE0104100) title "Nanoscale broadband spin-transfer-torque microwave detector" funded by Ministero degli Affari Esteri e della Cooperazione Internazionale (code CN16GR09,). The work at the Oakland University was in part supported by the National Science Foundation of the USA (EFMA-1641989, ECCS-1708982 ), by a grant from DARPA, and by a grant from the Center for NanoFerroic Devices (CNFD) and Nanoelectronics Research Initiative (NRI). The work at UCLA was supported by the Nanoscale Engineering Research Centre on Translational Applications of Nanoscale Multiferroic Systems (TANMS).

**Author contributions**

G. F., P. K. A. and Z. Z. initiated the work, designed the experiments, and coordinated all the activities. B. F. performed the experiments with support from J. A. K., S. L. and Y. F.; while M. C., R.T., and A.G performed the micromagnetic simulations to support the design of the MTJs and to reproduce the experimental data. G. F. and Z. Z. wrote the paper with contributions from P. K. A., A. S., and I. N. K. All authors contributed to the discussion and commented on the manuscript.

**Competing financial interests**

The authors declare no competing financial interests.

**Figures and figure captions:**

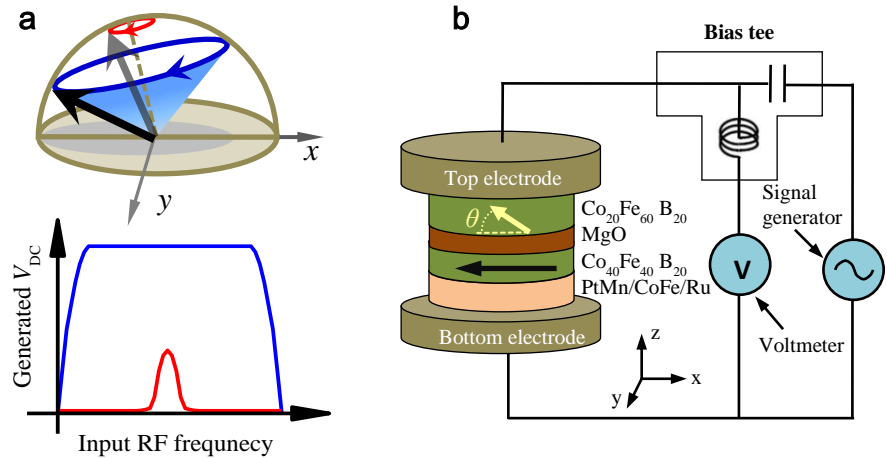

**Figure 1 | Schematic of the experiment and the measurement setup.** (**a**) Schematic image of a large out-of-plane precession in the nanomagnet. Under a RF current, the orbital centre of the free-layer magnetization rotates away from the initial state, causing a large change in the average resistance, producing a DC voltage. (**b**) Spintronic diode devices and circuit schematic used for diode effect measurements. The device is based on a magnetic tunnel junction with an in-plane magnetized reference layer and a perpendicularly magnetized free layer, separated by an MgO tunnel barrier. The generated voltage ($V_{DC}$) is measured by a Nanovolt meter in response to a RF current from a signal generator.

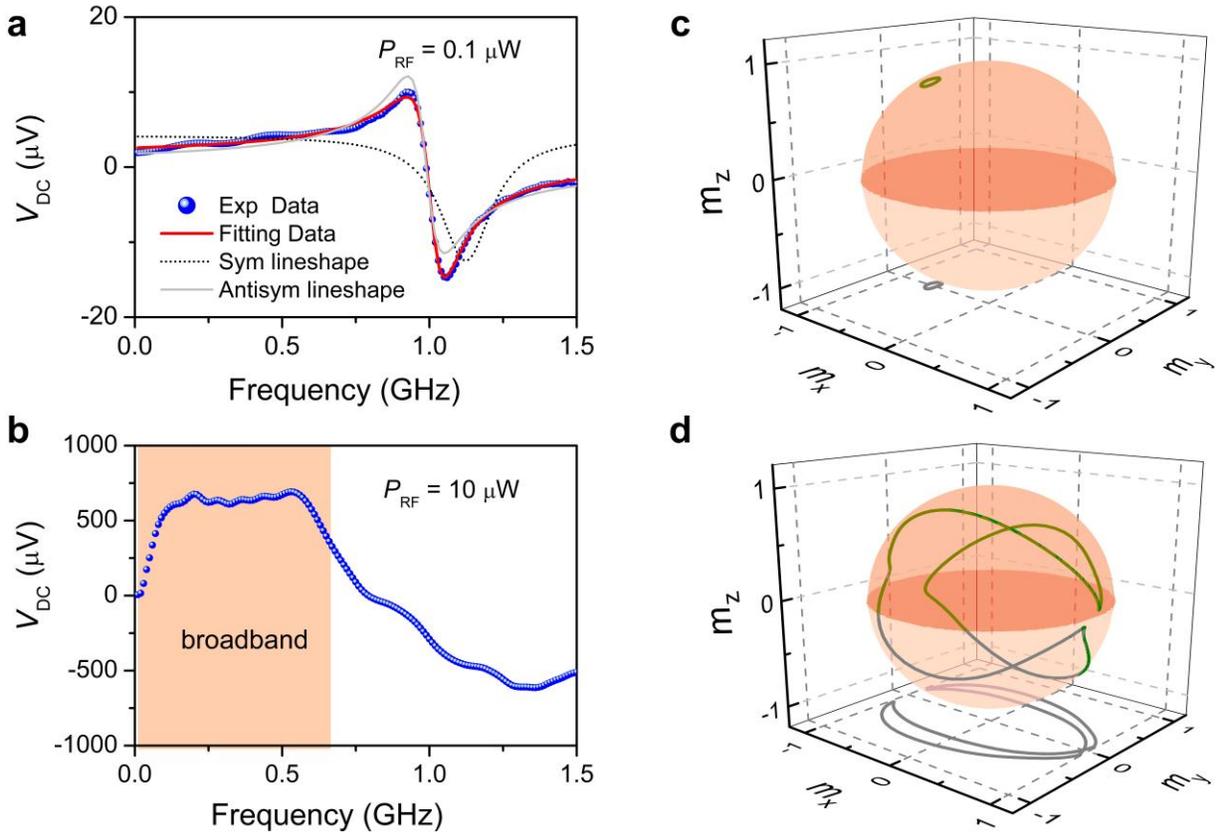

**Figure | 2 Performance of a spintronic nanogenerator device.** (**a** and **b**) Generated voltage ($V_{DC}$) as a function of RF frequency for RF powers ($P_{RF}$) of 0.1 μW and 10 μW. (**c** and **d**) The simulated trajectory of the free-layer magnetization at $P_{RF}$ = 0.1 μW and 10 μW, respectively.

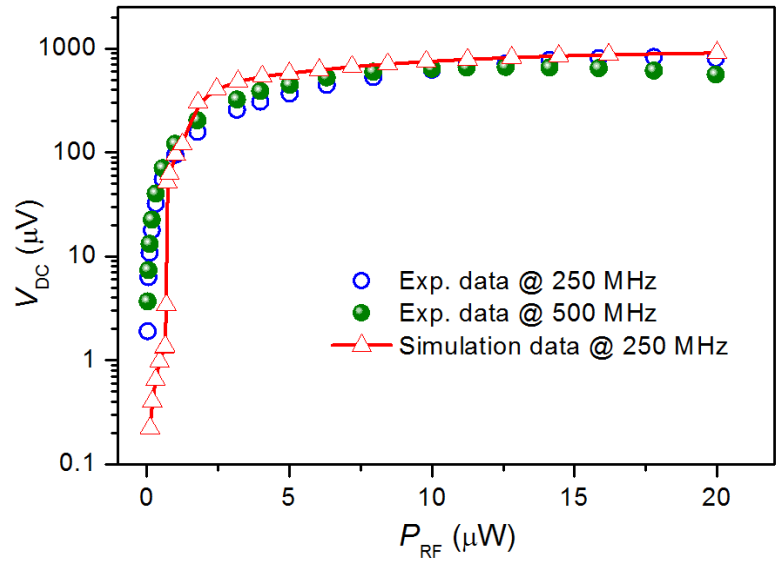

**Figure | 3 Comparison of experimental and simulated data.** The rectification voltage as a function of input power. The circles are the experimental data obtained at input frequencies $f_{ac}$ = 250 and 500 MHz, respectively. The triangles are the simulation data at an input frequency $f_{ac}$ = 250 MHz.

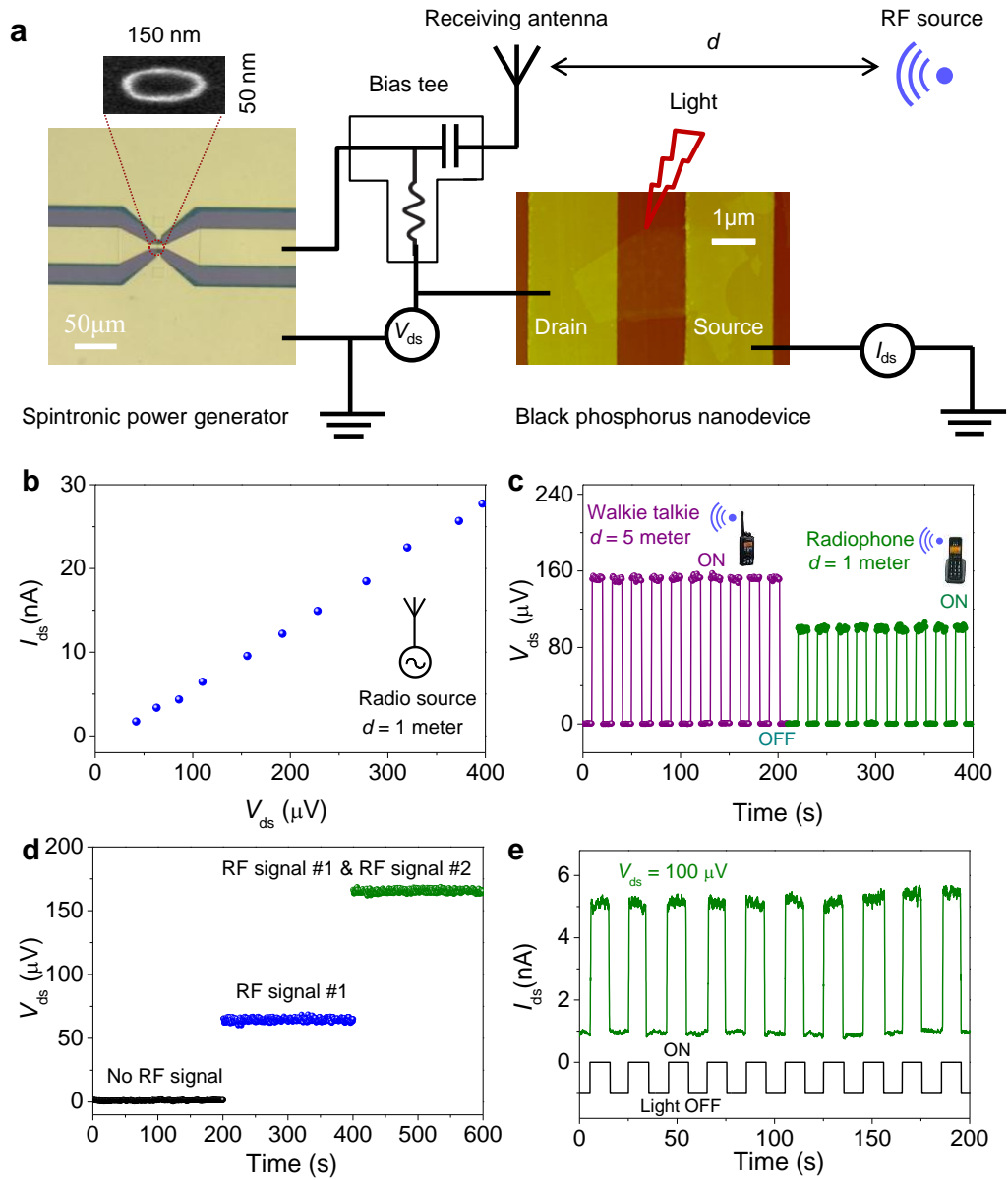

**Figure | 4 Demonstration of powering a nanodevice using spintronic power generator.** (**a**) Schematic diagram of the self-powered circuit: the drain-source supply for the BP nanodevice is provided by the nanoscale spintronic diode (NSD) power generator. The left shows the optical image of the spintronic generator with a SEM image of the magnetic tunnel junction nano-pillar, while the right shows the atomic force microscope image of the fabricated few-layer black phosphorus (BP) nanodevice and $d$ stands for the distance between the antenna and RF source. (**b**) Powering the drain-source terminal of the BP nanodevice with the NSD device (Sample **#S2** in

Supplementary Fig. 2) under RF energy (500 MHz) provided by signal generator. The graph shows the current-voltage characteristics. (**c**) Rectified d.c. voltages generated by RF signals that produced by Walkie talkie and Radiophone, respectively. (**d**) Rectified d.c. voltages as a function of time under the simultaneous action of two signals produced by the external RF sources: RF signal #1 (Signal generator) with a frequency of 650 MHz and a power of 3.2 µW; RF signal #2 (Radiophone) with a frequency of 923 MHz and a power of 4.0 µW. The data clearly demonstrates that the broad-band frequency response of the NSD device allows one to accumulate power from different RF energy sources. (**e**) Photoresponse of the BP transistor under light illumination at a voltage bias of 100 µV produced by the NSD that received the RF energy from radiophone.

Supplementary Information

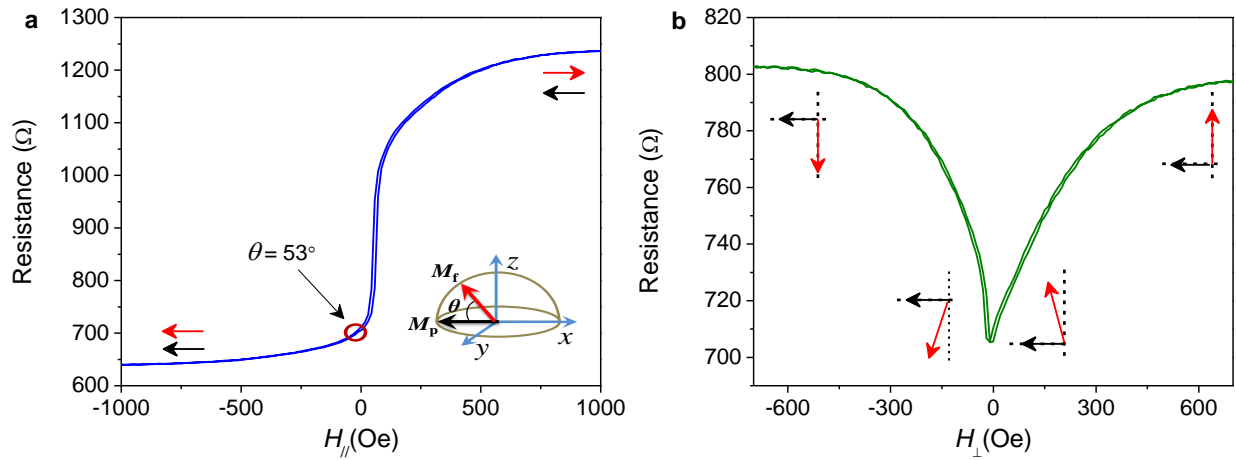

**Supplementary Figure 1: Magnetoresistance signal scans.** (**a** and **b**) Magnetoresistance-field curves of one typical magnetic tunnel junction devices under in-plane magnetic fields ($H_{//}$) and perpendicular magnetic fields ($H_{\perp}$) for $I_{dc}$ = 10 µA, respectively. The resistance scan as a function of the out-of-plane and in-plane fields indicates that the free layer is canted. The inset in (**a**) shows the Cartesian coordinate system and the direction of the magnetization vectors. The black (red) arrow denotes the magnetization direction of the reference (free) layer.

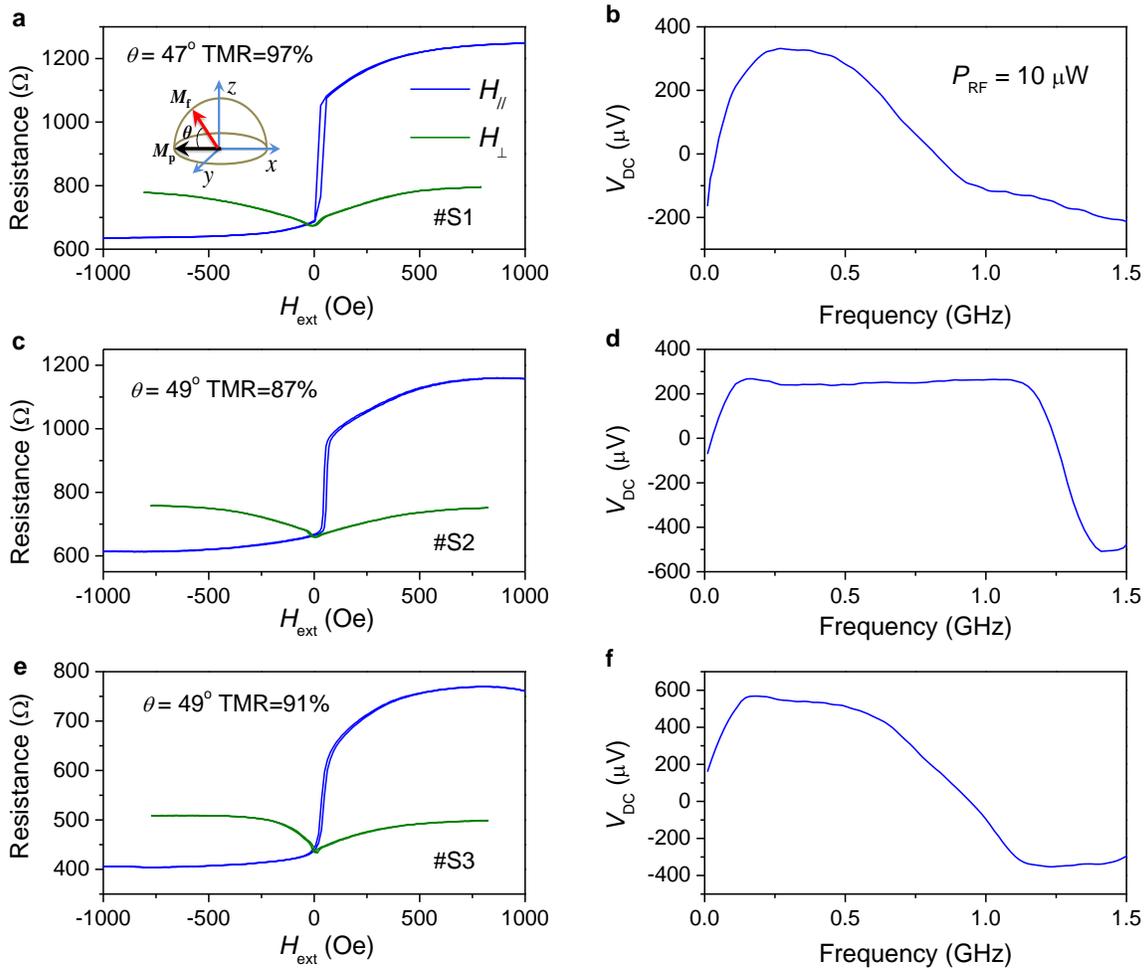

**Supplementary Figure 2: Experimental data from additional devices with canted free layer.** (**a**, **c** and **e**) The magnetoresistance curves of the magnetic tunnel junction devices (Sample #S1, S2 and S3) under in-plane magnetic field ($H_{//}$) and perpendicular magnetic field ($H_{\perp}$) for $I_{dc}=10$ μA. (**b, d** and **f**) Generated voltage ($V_{DC}$) as a function of RF frequency for an input RF power of 10 μW.

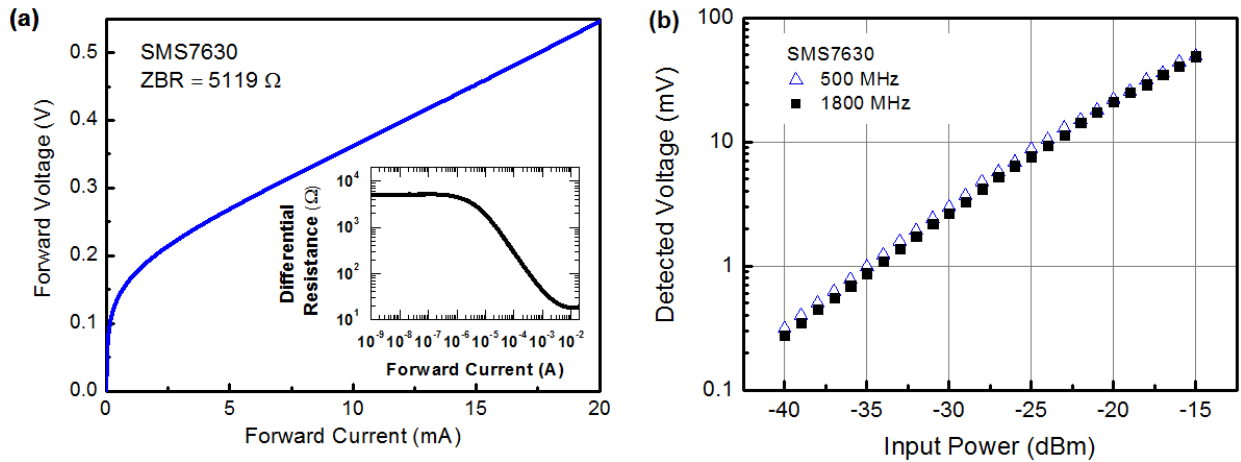

**Supplementary Figure 3: Properties of zero-bias SMS7630 Schottky diode (Manufacture "Skyworks").** (**a**) The *I-V* curves of one typical SMS7630 Schottky diodes. Inset in (**a**) is the differential resistance curves. (**b**) displays the corresponding rectifying behavior at different RF input powers at 500 MHz and 1800 MHz.

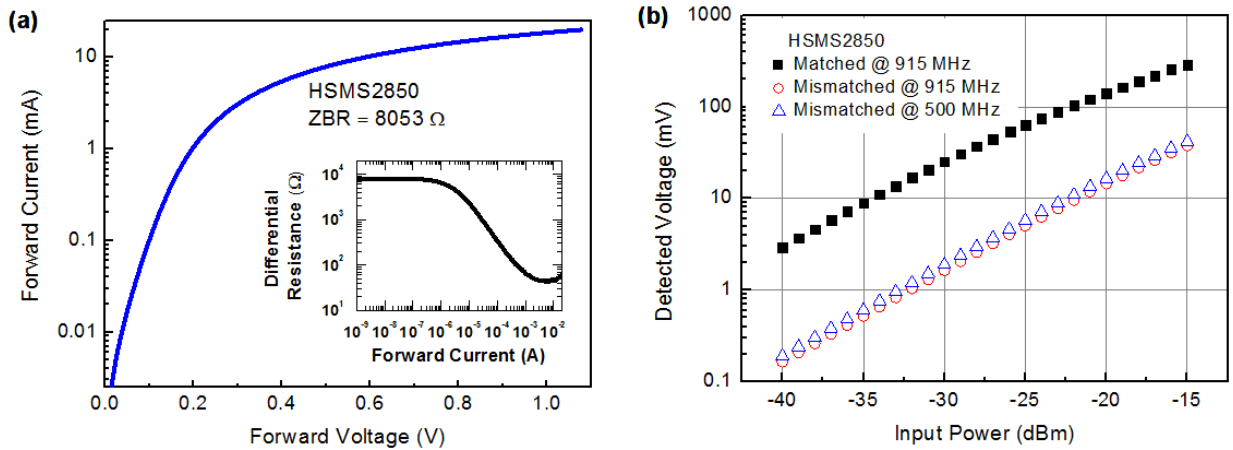

**Supplementary Figure 4: Properties of zero-bias HSMS2850 Schottky diode (Manufacture "Avago").** (**a**) The *I-V* curves of one typical HSMS2850 Schottky diodes. Inset in (**a**) is the differential resistance curves. (**b**) displays the corresponding rectifying behaviors at different RF input powers at 500 MHz and 915 MHz in the impedance matched condition (Measurement circuit is same as Figure 11 in Manufacture's datasheet[1] and impedance mismatched condition (same as Figure 1b in the main text).

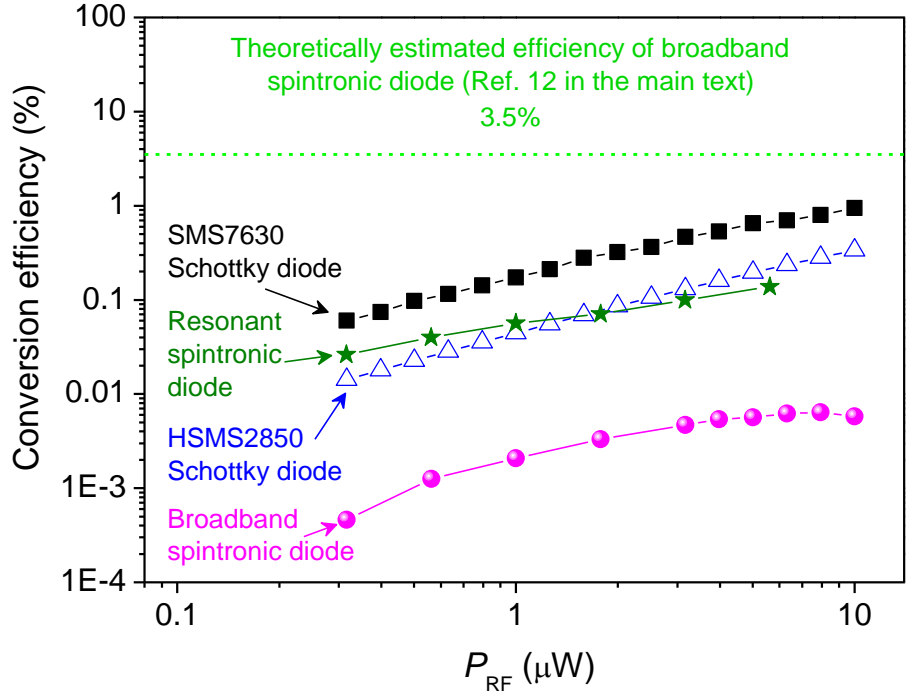

**Supplementary Figure 5: Comparison of RF-to-DC conversion efficiency.** The efficiencies in the Schottky diodes are calculated by the rectified DC voltages from Supplementary Fig. 3b and Fig. 4b at 500 MHz. The efficiency in the resonant FMR spintronic diode[2] is calculated from the rectified DC voltages at resonant frequency of 1.2 GHz. The efficiency of broadband spintronic diode (in this work) is calculated the rectified DC voltages at the 500 MHz in main text (Fig. 3). The theoretically estimated efficiency of broadband spintronic diode was adapted from Ref. 3.

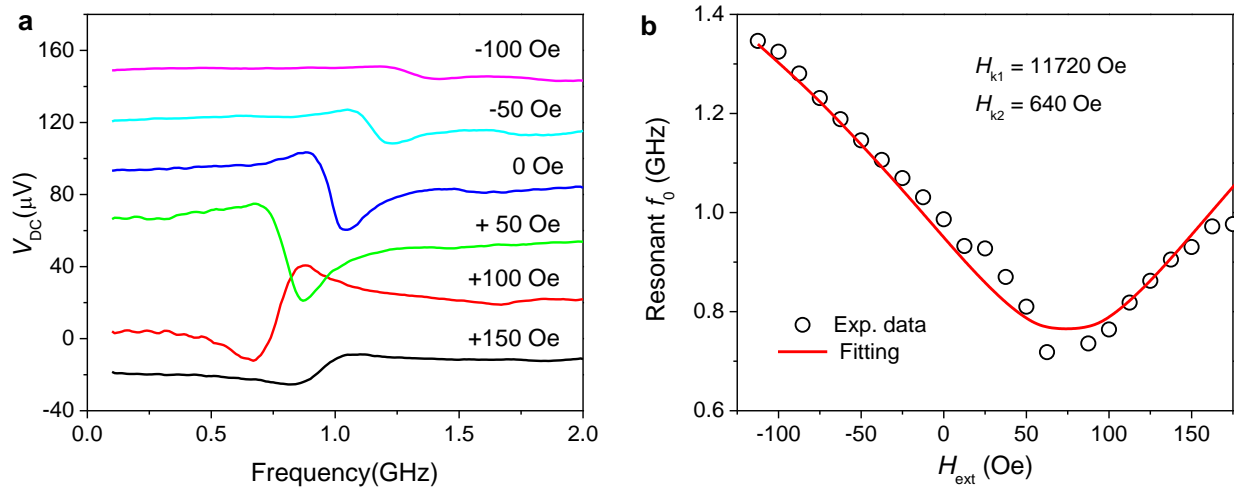

**Supplementary Figure 6:** Estimation of the perpendicular magnetic anisotropy (PMA) field. (**a**) Rectified voltage ($V_{DC}$) as a function of microwave frequency for various in-plane magnetic fields. (**b**) A comparison between the experimental data extracted from the panel (**a**) (black circles), and the analytical expression of the resonance frequency $f_0$ versus $H_{ext}$ (red line).

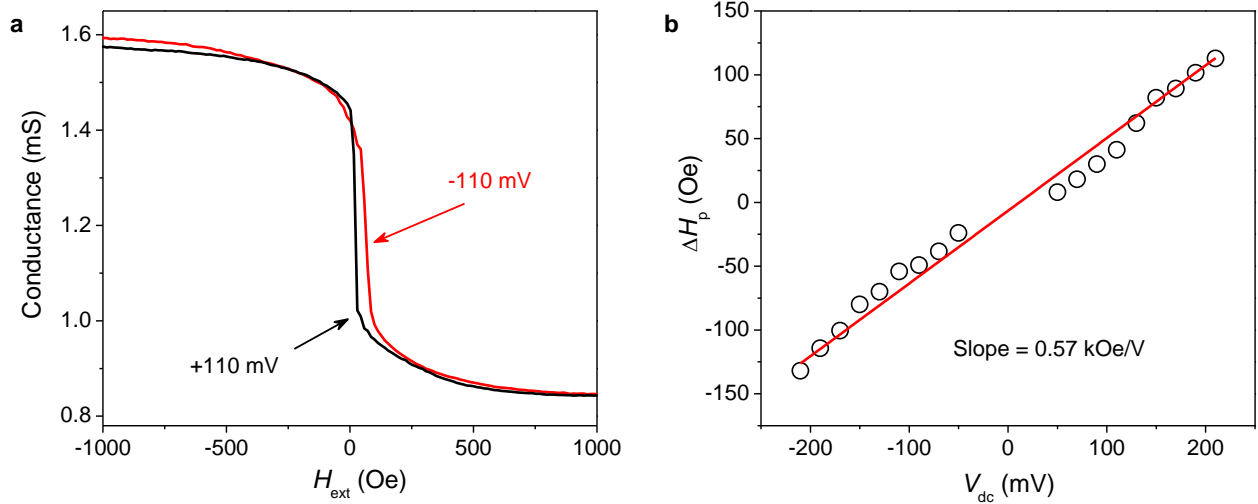

**Supplementary Figure 7:** Effect of the voltage on the perpendicular magnetic anisotropy. (**a**) Hysteresis loop of device conductance (G) versus in-plane magnetic field ($H_{ext}$) measured under ±110 mV direct voltage bias $V_{dc}$ for the device presented in the main text. (**b**) Variation of the effective perpendicular anisotropy field $\Delta H_p$ ($V_{dc}$) with d.c. voltage (black circles), extracted from the G($H_{ext}$) hysteresis loops via Equation (1) in Supplementary Ref. [4] and a linear fit to the data (red line).

**Supplementary Note 1. Experimental data from additional devices with canted free layer**

Here we present experimental data from the additional devices with canted free layer, showing that the broad-band behavior is robust. The devices (#S1, #S2 and #S3) whose data are presented below are fabricated on the same substrate as the device discussed in the main text. The junction resistance was measured as a function of the in-plane and out-of-plane applied magnetic field (Supplementary Fig. 2a, c and e). All those data confirms that a perpendicular anisotropy exists in the free layer, while the gradual approach to saturation for both field directions indicates the free-layer magnetization has a canted equilibrium state at zero bias field. The equilibrium angles ($\theta$) for S1, S2 and S3 at zero bias field are estimated to be 47°, 49° and 49°, respectively. The in-plane magnetoresistance ratios for S1, S2 and S3, defined as ($R_{AP}$-$R_P$)/$R_P$, where $R_{AP}$($R_P$) is the resistances in the antiparallel (parallel) configurations, are 97%, 87%, and 91%, respectively.

Supplementary Figure 2b, d and f display the rectified DC voltage as a function of the microwave frequency at $P_{RF}$ = 10 μW in the absence of magnetic field for S1, S2 and S3, respectively. It can be seen that all devices exhibit broad-band frequency response, similar to that discussed in the main text. These results confirm that the oblique design enables to achieve a broad-band frequency response at room temperature without bias magnetic field, which could be an efficient harvester of broadband ambient RF radiation.

**Supplementary Note 2. Estimation of the conversion efficiency for the spintronic diode and Schottky diodes**

We have performed an experiment to estimate the energy harvesting efficiency of Schottky diodes for comparison. The measurement setup is the same as the one that was previously used for spintronic diode measurements. We replaced our spintronic diode with Schottky diode. In our experiment, two types of Schottky diodes with lowest zero-bias resistances (SMS7630, manufactured by "**Skyworks**", and HSMS2850, manufactured by "**Avago**") were used. Supplementary Figure 3a displays the *I-V* curve of one typical SMS7630 Schottky diode. As it is emphasized by the differential junction resistance plotted in insets in Supplementary Fig. 3a, the SMS7630 diode has a zero-bias resistance (ZBR) of 5119 Ω. Supplementary Figure 3b shows the corresponding rectifying behaviors at different RF input powers, exhibiting a good rectifying behavior. For example, at an RF input power of 1 µW, the rectified DC voltage $V_{DC}$ is ~ 3.0 mV, consistent with the manufacture's datasheet[5].

Supplementary Figure 4a display the *I-V* curve of one HSMS2850 Schottky diodes from "Avago" company and the ZBR resistance is 8053 Ω. Good rectifying behavior was also observed. For example, at $P_{RF}$ = 1 µW, the rectified voltages for impedance matched condition are about 25 mV, which is consistent with that in the manufacture's datasheet[1]. Using our measurement setup (we refer as impedance mismatch condition), the rectified voltages are reduced to ~ 1.9 mV, which is in the same level as that in the SMS7630 diode.

For comparison, the rectified DC voltages obtained at a frequency of 500 MHz and the maximum rectified DC voltages of each Schottky diode and broadband spintronic diode were used. Supplementary Fig. 5 summarizes the calculated efficiencies for different diodes. It can be seen that, in the current stage, the broadband spintronic diodes has small efficiency than the Schottky diodes (SMS7630 and HSMS2850) with best low-power capability. For example, at $P_{RF}$ = 3.2 µW, the spintronic diode has an efficiency of 0.005%, compared to 0.4% for the SMS7630 Schottky diodes and 0.13% for the HSMS2850 Schottky diodes. Despite the fact that the broadband spintronic diode in the present stage doesn't provide a satisfying efficiency, an improved diode design could lead to a significant improvement of the efficiency:

1) **Optimization of the resistance oscillation $\Delta R_s$.** It is important to note that the generated DC voltage in this work is a small fraction of the maximum value. The maximum $V_{DC}$ is determined by the resistance oscillation $\Delta R_s$ ($V_{DC}^{max} \approx \frac{1}{\sqrt{2}} I_{RF} \Delta R_s$, where $I_{RF}$ is the input RF current in the MTJ), which is dependent on the magnetoresistance (TMR) effect, i.e. $\Delta R_{s-max} = (R_{AP} - R_P)/2$, where the resistances in the antiparallel ($R_{AP}$) and parallel ($R_P$) configurations, respectively. Hence, the output DC voltage is limited by the maximum MR effect at a given bias current. For example, for the device presented in this work, $R_{AP}$ = 1236 Ω, $R_P$ = 640 Ω, at $I_{RF}$ = 60 µA ($P_{RF}$ =10 µW), the measured $V_{DC}$ is 0.65 mV, while the expected maximum $V_{DC}^{max}$ can be as high as 12.6 mV (namely, $V_{DC}$ ~ 5% $V_{DC}^{max}$). Therefore, the efficiency corresponding to the $V_{DC}^{max}$ values is significantly enhanced, up to 1.1%.

2) **Optimizing the magnetic stack of the spintronic diode to enhance the TMR effect.**
   In the presented spintronic diode, the TMR ratio is 93%, which promises a high potential for improvement in the future. For examples, Ikeda *et al* reported a TMR ratio of 604% in an in-plane magnetized MTJ structure CoFeB/MgO/CoFeB at room temperature[6]. Almasi *et al.* obtained a TMR as high as 208% in a perpendicular magnetic tunnel junction[7]. Kubota *et al.* theoretically predicted a TMR of 600% for the $D0_{22}$-$Mn_3Ga$/MgO/$Mn_3Ga$ (001) MTJ system[8]. Larger TMR effect will result in larger efficiency.

3) In addition, the sensitivity could be enhanced by optimizing the nonlinearity of the MTJ characteristics[9], or by taking advantage of the easy plane anisotropy as predicted by micromagnetic simulations that we have performed.

Our previous work exhibiting a resonant ferromagnetic resonance rectifying behavior reported a larger sensitivity of about 1000 mV/mW at zero DC bias[2]. At ferromagnetic resonance, the efficiency is significantly enhanced, which is comparable to those in the Schottky diodes as shown in Supplementary Fig. 5 (please see the olive stars). For example, at $P_{RF}$ = 3.2 µW, the spintronic diode has an efficiency of 0.1%, which is close to that HSMS2850 Schottky diodes. Therefore, following the projection of theoretical work[3], the spintronic diodes with optimized structures (e.g. large TMR ratio) are expected to have better efficiency than that in the current Schottky diodes.

**Supplementary Note 3. Estimation of the perpendicular magnetic anisotropy (PMA) field**

In order to evaluate the PMA in the magnetic tunnel junction (MTJ), we performed the spin-torque ferromagnetic resonance (ST-FMR) measurement under various external in-plane magnetic fields at zero d.c. bias voltage. Supplementary Figure 6a shows the rectified D.C. voltages of the device presented in the main text as a function of the microwave frequency for different in-plane magnetic fields at $P_{RF}$ = 0.32 µW. The ST-FMR spectra can be well fit by a sum of symmetric and antisymmetric Lorentzians with identical resonance frequencies $f_0$ (Supplementary Ref.[4]). The first- ($H_{k1}$) and second-order ($H_{k2}$) perpendicular anisotropy field can be deduced from the magnetic field dependence of the resonance frequency $f_0$ (Supplementary

Ref.[4]). For the device presented in the main text, the values of $H_{k1}$ and $H_{k2}$ are estimated to be 11720 Oe and 640 Oe, respectively (Supplementary Fig. 6b).

**Supplementary References:**